# Cryptanalysis of an efficient signcryption scheme with forward secrecy based on elliptic curve [†]

Mohsen Toorani [‡]         Ali A. Beheshti

## Abstract

*The signcryption is a relatively new cryptographic technique that is supposed to fulfill the functionalities of encryption and digital signature in a single logical step. Several signcryption schemes are proposed throughout the years, each of them having its own problems and limitations. In this paper, the security of a recent signcryption scheme, i.e. Hwang et al.'s scheme is analyzed, and it is proved that it involves several security flaws and shortcomings. Several devastating attacks are also introduced to the mentioned scheme whereby it fails all the desired and essential security attributes of a signcryption scheme.*

## 1. Introduction

The encryption and digital signature are two fundamental cryptographic tools that can guarantee the unforgeability, integrity, and confidentiality of communications. Until the before decade, they have been viewed as important but distinct building blocks of various cryptographic systems. In public key schemes, a traditional method is to digitally sign a message then followed by an encryption (signature-then-encryption) that can have two problems: Low efficiency and high cost of such summation, and the case that any arbitrary scheme cannot guarantee the security. The signcryption is a relatively new cryptographic technique that is supposed to fulfill the functionalities of digital signature and encryption in a single logical step and can effectively decrease the computational costs and communication overheads in comparison with the traditional signature-then-encryption schemes. The first signcryption scheme was introduced by Zheng in 1997 [1] but it fails the forward secrecy of message confidentiality [2]. Zheng also proposed an elliptic curve-based signcryption scheme that saves 58% of computational and 40% of communication costs when it is compared with the traditional elliptic curve-based signature-then-encryption schemes [3]. There are also many other signcryption schemes that are proposed throughout the years, each of them having its own problems and limitations, while they are offering different level of security services and computational costs.

In a signcryption scheme, the sender usually uses the public key of recipient for deriving a session key of a symmetric encryption, while the recipient uses his private key for deriving the same session key. Exposure of session keys can be a devastating attack to a cryptosystem since such an attack typically implies that all the security guarantees are lost. In this paper, we show that a recent signcryption scheme, i.e. Hwang et al.'s scheme [4] have such vulnerability and many other security flaws and shortcomings that are described throughout the paper. For the case of brevity, in this paper we refer to Hwang et al.'s scheme [4] as HLS. This paper is organized as follows. Section 2 briefly presents some preliminaries on signcryption and its desired attributes. Section 3 is devoted to security analysis of the HLS signcryption scheme, and Section 4 provides the conclusions.

## 2. Preliminaries on signcryption

A signcryption scheme $\Sigma = (Gen, SC, USC)$ typically consists of three algorithms: Key Generation (*Gen*), Signcryption (*SC*), and Unsigncryption (*USC*). *Gen* generates a pair of keys for any user *U*: $(SDK_U, VEK_U) \leftarrow Gen(U, \lambda)$ where $\lambda$ is the security parameter, $SDK_U$ is the private signing/decryption key of user $U$, and $VEK_U$ is his public verification/encryption key. For any message $m \in M$, the signcrypted text $\sigma$ is obtained as $\sigma \leftarrow SC(m, SDK_S, VEK_R)$ where $S$ denotes the sender, and $R$ is the recipient. $SC$ is generally a probabilistic algorithm while $USC$ is most likely to be deterministic where $m \cup \{\bot\} \leftarrow USC(\sigma, SDK_R, VEK_S)$ in which $\bot$ denotes the invalid result of unsigncryption. A formal proof for the security of signcryption is provided in [5]. Any signcryption scheme should have the following properties:

1) **Correctness:** A signcryption scheme is correct only if for any sender $S$, recipient $R$, and message $m \in M$, $\exists USC(SC(m, SDK_S, VEK_R), SDK_R, VEK_S) = m$.

2) **Efficiency:** The computational costs and communication overheads of a signcryption scheme should be smaller than those of the best known signature-then-encryption schemes with the same provided functionalities.

3) **Security:** A signcryption scheme should simultaneously fulfill the security attributes of an encryption scheme and those of a digital signature. Such additional properties mainly include: *Confidentiality*, *Unforgeability*, *Integrity*, and *Non-repudiation*. Some signcryption schemes provide further attributes such as *Public verifiability* and *Forward secrecy of message confidentiality* while the others do not provide them. Such properties are the attributes that are required in many applications while the others may not require them. Hereunder, the above-mentioned attributes are briefly described.

- **Confidentiality:** It should be computationally infeasible for an adaptive attacker to gain any partial information on the contents of a signcrypted text, without knowledge of the sender's or designated recipient's private key.
- **Unforgeability:** It should be computationally infeasible for an adaptive attacker to masquerade an honest sender in creating an authentic signcrypted text that can be accepted by the unsigncryption algorithm.
- **Non-repudiation:** The recipient should have the ability to prove to a third party (e.g. a judge) that the sender has sent the signcrypted text. This ensures that the sender cannot deny his previously signcrypted texts.
- **Integrity:** The recipient should be able to verify that the received message is the original one that was sent by the sender.
- **Public Verifiability:** Any third party without any need for the private key of sender or recipient can verify that the signcrypted text is the valid signcryption of its corresponding message.
- **Forward Secrecy of message confidentiality:** If the long-term private key of the sender is compromised, no one should be able to extract the plaintext of previously signcrypted texts. In a regular signcryption scheme, when the long-term private key is compromised, all the previously issued signatures will not be trustworthy any more. Since the threat of key exposure is becoming more acute as the cryptographic computations are performed more frequently on poorly protected devices such as mobile phones, the forward secrecy seems an essential attribute in such systems.

---

[‡] Corresponding Author, ResearcherID: A-9528-2009

Many of the proposed signcryption schemes include modular exponentiation while some of them are based on elliptic curves. The elliptic curve-based solutions are usually based on difficulty of *Elliptic Curve Discrete Logarithm Problem* (ECDLP) that is computationally infeasible under certain circumstances [6]. The elliptic curve-based systems can attain to a desired security level with significantly smaller keys than that of required by their exponential-based counterparts. As an example, it is believed that a 160-bit key in an elliptic curve-based system provides the same level of security as that of a 1024-bit key in an RSA-based system [6]. This can enhance the speed and leads to efficient use of power, bandwidth, and storage that are the basic limitations of resource restricted devices. Throughout this paper, *Alice* is the sender or the initiator of the protocol, *Bob* is the designated recipient, and *Mallory* is the malicious active attacker.

## 3. Cryptanalysis of the HLS Signcryption Scheme

A schema of the HLS signcryption scheme [4] is depicted in Figure 1 where the deployed notations are described in Figure 2. The public keys of *Alice* and *Bob* in HLS are generated as $U_A = d_A G$ and $U_B = d_B G$. The HLS claims to provide the attributes of *confidentiality*, *unforgeability*, *integrity*, *public verifiability*, and *forward secrecy of message confidentiality*. However, as we prove in this section, it has several security flaws so that it fails all the desired and essential security attributes of a signcryption scheme.

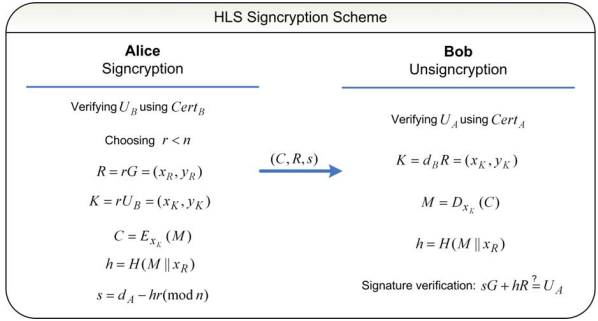

**Figure 1. Hwang et al.'s Signcryption Scheme (HLS) [4]**

| Notations Guide | | | |
|---|---|---|---|
| $\in_R$ | Chosen randomly | $O$ | Point of elliptic curve at infinity |
| $M$ | Plaintext | $n$ | Order of $G$ ($nG=O$) |
| $C$ | Ciphertext | $x_R$ | x-coordinate value of point R |
| $s$ | Digital signature | $y_R$ | y-coordinate value of point R |
| $H$ | One-way hash function | $d_A/U_A$ | Private/Public key of *Alice* |
| $\|\|$ | Concatenation | $d_B/U_B$ | Private/Public key of *Bob* |
| $G$ | Base point of elliptic curve | $E_k(.)/D_k(.)$ | Symmetric encryption/decryption with the secret key $k$ |

**Figure 2. Deployed notations**

1) The security of HLS completely depends on the secrecy of random number $r$. It does not have any resilience to disclosure of such ephemeral parameter, and the long-term private key of initiator (*Alice*) will be simply divulged with disclosure of $r$. The point $R$ is obtained as $R = rG$, and it is clearly sent to *Bob*. It does not differ for the users that have selected the same value of $r$. If *Mallory* knows the corresponding $r$ of $R$, he can easily deduce the static private key of *Alice* from an intercepted pair of $(C,R,s)$. He calculates $K = rU_B = (x_K, y_K)$, finds the secret key of encryption ($k = x_K$), decrypts the ciphertext as $M = D_k(C)$, and calculates $h = H(M \| x_R)$. He then simply deduces the long-term private key of *Alice* as $d_A = (s + hr) \mod n$. Therefore, the confidentiality, unforgeability, non-repudiation, and other claimed security attributes of the HLS completely depend on the secrecy of $r$ and will be completely failed with its disclosure. This attack is feasible due to the weakness in session key derivation function of the HLS.

Although it is believed that finding the corresponding $r$ of $R$ is in deposit of solving the ECDLP, there are some practical circumstances where this is not the case. Hereunder, we describe two practical situations in which the adversary can find the corresponding $r$ of some $R$ without solving the ECDLP. The first feasibility is that many applications may boost their performance by pre-computing the ephemeral pairs of $(r, R)$ for their later uses. This may be applied to low-power devices as well as high volume servers. In this case, the stored pairs are more vulnerable to leakage than the long-term private keys. The latter may be stored on a hardware protected storage media while the former is typically stored on disks and hence is exposed to more vulnerability. If an adversary could have any access to such stored pairs, he can easily deduce the long-term private key of the sender by following the above-mentioned approach. The second feasibility is to misuse the possible weaknesses of the deployed random number generators. The generated random numbers are actually pseudo-random and may have some biases, especially when they are generated in resource restricted devices. *Mallory* runs the deployed random number generator of his victims, produces the most probable pairs of $(r, R)$, and saves them offline. He then intercepts the *Bob*'s terminal that would have many transactions everyday. He observes the clearly sent $R$ in the intercepted messages and picks those messages that he has their $R$ in his compiled list. He simply deduces the long-term private keys of the corresponding senders from such chosen signcrypted texts by following the previously stated method. This can be considered as a *chosen-ciphertext attack*. He uses the deduced private keys for impersonating himself as the legitimate users and performing his malicious activities. Although $r$ is randomly selected, it is feasible that some users produce the same random number at the same time so *Mallory* can easily deduce the long-term private keys of all the senders that have the same value of $R$ in their signcrypted texts provided that he knows the corresponding $r$ of that $R$. If *Mallory* aims a definite entity, he may wait until his victim sends an $R$ that he has it in his compiled list. Until then, *Mallory* can make his list richer.

Although the mentioned attack works for awkward implementations of the HLS, it is completely regarded to its weak session key derivation function that includes a simple elliptic curve point multiplication and taking the x-coordinate of the product as the session key.

2) Although HLS considers verifying public keys using the certificates, the certificate validation is not considered itself. The process of certificate validation includes [7]:
(a) Verifying the integrity and authenticity of the certificate by verifying the CA's signature on the certificate.

(b) Verifying that the certificate is not expired.
(c) Verifying that the certificate is not revoked.
Without a certificate validation, the certificates and public keys can be easily forged.

3) HLS does not consider the public key validation so it is feasible to get certificates for the invalid public keys. An invalid public key is of a small order resided on an invalid-curve, and can be misused for an *invalid-curve attack* [6, 8]. The selected elliptic curve of HLS has the Weierstrass equation of the form $y^2 = x^3 + ax + b$ defined over the finite field $F_q$ so its corresponding invalid-curve has the form of $y^2 = x^3 + ax + b'$. The public key of user $U$, $U_U = (x_{U_U}, y_{U_U})$ is valid if all the following conditions are simultaneously satisfied [8]:
(a) $U_U \neq O$.
(b) $x_{U_U}$ and $y_{U_U}$ should have the proper format of $F_q$ elements.
(c) $U_U$ should satisfy the defining equation of *E*.

Traditionally, the public key validation is not considered in the PKI standards (such as [9] and [10]), and the *Certificate Authority* (*CA*) just performs a proof of possession by checking the user's signature over a message of a predetermined format so it is feasible to get a certificate for an invalid public key if the public key validation is not considered. Antipa et al. [8] demonstrated how to get a certificate for an invalid public key when *CA* uses the ECDSA for its digital signatures. In the HLS, *CA* does not verify whether each entity really possesses the corresponding private key of its claimed public key or not. Such shortcoming exposes it to the mentioned vulnerability.

4) The delivery confirmation or a receipt from the recipient is necessary for some applications. Although HLS is one-pass, the implementer may add a confirmation step in which *Bob* sends *Alice* a confirmation message perhaps in addition to a *Message Authentication Code* (MAC) in which the session key of encryption is used as the key. Since the validity verification of ephemeral public key (point *R*) is not included in unsigncryption phase of the HLS, it can be misused for an *invalid-curve attack* [8] whereby *Alice* is capable of deducing the long-term private key of *Bob*. Here is how the attack works [8]. *Alice* chooses an invalid-curve containing a point $W_i$ of small order $g_i$. She uses $W_i$ instead of *R*, proceeds the signcryption, and sends $(W_i, C, s)$ to *Bob*. In the unsigncryption, *Bob* computes $K = d_B W_i = (x_K, y_K)$, and proceeds the unsigncryption. Finally, when *Bob* sends the confirmation message $M'$ and its corresponding tag $z = MAC_{x_K}(M')$ to *Alice*, she can easily determine a point $K' \in <W_i>$ satisfying $z = MAC_{x_{K'}}(M')$ due to the small order of her chosen $W_i$ point. Hence, *Alice* can find $d_{g_i}^2 \equiv d_B^2$ with $\frac{g_i}{2}$ number of trials. *Alice* selects other $W_i$ points of different orders $g_i$, and repeats the above-mentioned attack. The order of such chosen $W_i$ points should be relatively prime, i.e. we should have $\gcd(g_i, g_j) = 1, \quad \forall i \neq j$. Such points can be selected from different invalid-curves. Each round of attack gives $d_{g_i}^2 \equiv d_B^2$. Ultimately, *Alice* finds the private key of *Bob* using the *Chinese Remainder Theorem* (CRT) [7] while *Bob* is unaware that such an attack is taking place.

5) The HLS is vulnerable to the *Unknown Key-Share* (UKS) attack. In an UKS attack, two parties compute the same session key but have different views of their peers in the key exchange. In an UKS attack, an adversary interferes with the *Alice*'s and *Bob*'s communication so that *Alice* correctly believes that her session key is shared with *Bob* while *Bob* mistakenly believes that the session key is shared with another entity. This can be accomplished whenever *Mallory* can convince one of the honest parties that he has the knowledge of the session key. Further issues on the practical attack scenarios and significance of the UKS attack are provided in [11]. The UKS attack is feasible when a key exchange protocol fails to provide an authenticated binding between the session key and identifiers of the honest entities. Since the private key and identifier of *Alice* is not involved in the session key derivation function of the HLS, it does not have any resilience to the UKS attack.

6) The HLS does not really have its claimed attribute of *forward secrecy of message confidentiality*. The outsider and insider security are two notions of security that are usually considered in signcryption schemes. While the outsider security assumes that the adversary is neither sender nor the recipient, the insider security allows an adversary to be sender or recipient. The *forward secrecy of message confidentiality* is an attribute that is provided through the insider security. HLS claims to have such an attribute because its message confidentiality relies on two secret factors: the long-term private key of *Alice*

($d_A$), and the ephemeral random number *r*. However, anyone who has $d_B$ can simply recover the signcrypted text and deduce the corresponding random number *r* as $r = (d_A - s)h^{-1} \mod n$. When $d_A$ is revealed, *Mallory* who could obtain $d_A$, may also request *Bob* to compute the corresponding *r* for him so he can simply recover the signcrypted text without any need for the knowledge of $d_B$. Regardless of its practical benefits, this invalidates the definition of *forward secrecy of message confidentiality*.

7) There is not any provision for the *key control* in the HLS so the plaintext may be encrypted with a weak or even a full-zero key. There is also no checking for $K \neq O$.

8) Domain parameters of the HLS are not exactly selected. Practically, there are other considerations that should be taken into account when selecting the domain parameters in order to thwart several potential attacks to elliptic curves. Such requirements are not considered in domain parameters of the HLS, and can make it vulnerable to several kinds of attacks if the implementer unconsciously selects the domain parameters in the range of such non-stated conditions. In order to thwart the *small subgroup attacks* [6], the point *G* should be of a prime order *n* and we should have $n > 4\sqrt{q}$ [12]. The condition $n > 4\sqrt{q}$ that is not considered in the HLS, can make it vulnerable to *small subgroup attacks*. Furthermore, to protect against other known attacks on special classes of elliptic curves, *n* should not divide $q^i - 1$ for some integer *i*, $n \neq q$ should be satisfied, and the curve should be non-supersingular [12]. Such considerations are not also considered in specifications of the HLS's domain parameters.

9) The zero value is not excluded from the variation range of *r*. There is also no checking for $R \neq O$ in signcryption phase of the HLS. The signature generation function is not well-designed so if $r = 0$ occurs, the clearly sent digital signature exhibits the long-term private key of *Alice* ($s = d_A$).

## 4. Conclusions

The security of Hwang et al.'s signcryption scheme (HLS) [4] is evaluated in this paper, and it is proved that it involves several security flaws and shortcomings as it fails all the desired and essential security attributes of a signcryption scheme. As it is illustrated in this paper, one of the most important security vulnerabilities of the HLS is due to its weak session key establishment that can lead to disclosure of long-term private key of the initiator, and the case that it does not consider many essential considerations such as validity verification of static and ephemeral public keys and certificates. There are also other shortcomings that were explained throughout the paper.